\documentclass[graybox]{svmult}

\usepackage{amsmath,amssymb}
\usepackage{mathptmx}       
\usepackage{helvet}         
\usepackage{courier}        
\usepackage{type1cm}        
\usepackage{makeidx}         
\usepackage{graphicx}        
\usepackage{multicol}        
\usepackage{multirow}
\usepackage[bottom]{footmisc}
\usepackage{natbib}
\usepackage{hyperref}
\usepackage[usenames,dvipsnames]{color}
\usepackage[normalem]{ulem}

\usepackage{enumerate}
\renewcommand{\d}{\mathrm{d}}
\renewcommand\arraystretch{1.2}

\newcommand{\bi}{\begin{itemize}}
\newcommand{\ei}{\end{itemize}}
\newcommand{\be}{\begin{equation}}
\newcommand{\ee}{\end{equation}}
\newcommand{\beq}{\begin{equation}}
\newcommand{\eeq}{\end{equation}}
\newcommand{\bea}{\begin{eqnarray}}
\newcommand{\eea}{\end{eqnarray}}
\newcommand{\bqu}{\begin{quote}}
\newcommand{\equ}{\end{quote}}
\newcommand{\bctr}{\begin{center}}
\newcommand{\ectr}{\end{center}}
\newcommand{\bd}{\begin{description}}
\newcommand{\ed}{\end{description}}
\newcommand{\bdm}{\begin{displaymath}}
\newcommand{\edm}{\end{displaymath}}
\newcommand{\lsim}{\mbox{$\:\stackrel{<}{_{\sim}}\:$} }

\newcommand{\om}{\rm \Omega_{\rm m}}

\newcommand{\oll}{\rm \Omega_\Lambda}

\newcommand{\ok}{\rm \Omega_{\rm k}}

\newcommand{\kms}{\ensuremath{{\rm km\, s}^{-1}}} 
\newcommand{\kmsmpc}{\ensuremath{{\rm km\, s}^{-1}{\rm Mpc}^{-1}}}

\include{jdf}

\begin{document}

\title*{Characterising Dark Energy through supernovae\\
{\small Chapter for publication in the Handbook of Supernovae, Springer, edited by Athem W. Alsabti and Paul Murdin.}
}
\titlerunning{Characterising Dark Energy}
\author{Tamara M.\ Davis and David Parkinson}
\authorrunning{Tamara Davis \& David Parkinson}
\institute{School of Mathematics and Physics, University of Queensland, QLD 4072; {tamarad@physics.uq.edu.au}} 
%
%
\maketitle

\abstract{
Type Ia supernovae are a powerful cosmological probe, that gave the first strong evidence that the expansion of the universe is accelerating.  Here we provide an overview of how supernovae can go further to reveal information about what is causing the acceleration, be it dark energy or some modification to our laws of gravity.  We first summarise the many different approaches used to explain or test the acceleration, including parametric models (like the standard model, $\Lambda$CDM), non-parametric models, dark fluid models such as quintessence, and extensions to standard gravity. We also show how supernova data can be used beyond the Hubble diagram, to give information on gravitational lensing and peculiar velocities that can be used to distinguish between models that predict the same expansion history.  Finally, we review the methods of statistical inference that are commonly used, making a point of separating parameter estimation from model selection. }

\section{Introduction}
\label{sec:intro}

Discovering that the universe is accelerating is one thing.  Determining what is causing the acceleration is quite another.  Ever since type Ia supernovae confirmed \citep{riess98,perlmutter99} earlier hints \citep{efstathiou90,ostriker95,krauss95,yoshii95}  that the universe was accelerating, cosmologists have focussed on determining why.  Whatever the cause, we give it the name {\em dark energy}\index{dark energy}.

{\em Dark energy} covers a wide-range of possible explanations, from the energy of the vacuum through to beyond-standard theories of gravity.  
In very general terms one can split explanations of the acceleration into those that modify gravity, i.e. change the $G_{\mu\nu}$ term in the Einstein equation\index{Einstein equation}, 
\beq G_{\mu\nu} = 8\pi T_{\mu\nu}, \eeq
and those that modify the contents of the universe, i.e. change the $T_{\mu\nu}$ term.   In some cases the distinction is operationally irrelevant.  For example, the cosmological constant\index{cosmological constant}, introduced by Einstein such that $G_{\mu\nu}\rightarrow G_{\mu\nu}+g_{\mu\nu}\Lambda$, is exactly equivalent to adding a vacuum energy\index{vacuum energy} to $T_{\mu\nu}$, where the cosmological constant is related to the energy density of the vacuum $\Lambda=8\pi\rho_{\rm vac}$.   Some researchers use  {\em modified gravity}\index{modified gravity} to distinguish explanations that change the theory of gravity from those that introduce a new component to the stress-energy-momentum tensor of general relativity ($T_{\mu\nu}$), but here we will stick to {\em dark energy} to generically cover all possibilities.

In this chapter we aim to equip you with the tools you require to test new cosmological models against supernova data. To that end we review the techniques available to determine which is the best model, an aim that  differs from finding the best-fit parameters within a model (Sect.~\ref{sect:stats}). We then summarise the range of models currently under consideration (and some classics that have already been ruled out; Sect.~\ref{sect:models}).  We conclude by looking at how supernovae are being used to test cosmology through the inhomogeneities in the Hubble diagram caused by peculiar velocities and gravitational lensing (Sect.~\ref{sec:beyondhubble}).

\section{Statistical Inference - Model testing vs parameter fitting}\label{sect:stats}

The supernovae data, like any measurement of cosmological distances, cannot be used to determine the properties of the Universe directly. In this sense the Universe cannot be `weighed' in the manner of an object or particle on Earth, since the properties of dark energy and dark matter, or the nature of the acceleration, are classic examples of an {\bf Inverse Problem}\index{Inverse Problem} ---  we start with a set of results (in this case the magnitudes of distant supernovae),  then compare them with predictions from some assumed causes. This is opposite to a forward problem, which starts with a set of causes, and then computes the results.

Statistical inference of this type is commonly done in a {\bf Bayesian} framework, where the probabilities associated with causes as well as results can be evaluated. Bayesian statistics\index{Bayesian statistics} provides two complementary tools:
\begin{itemize}
\item {\bf Parameter fitting}\index{Parameter fitting}, where the data is used to evaluate the probabilities associated with the parameters of a given assumed model.
\item {\bf Model selection}\index{Model selection}, where the data is used to choose between different models or explanations of the data.
\end{itemize}
To determine which model  best explains dark energy we need to test each model against data and rank them based on some model selection statistic.    Rather than trying to find the best fitting parameters within a model, we are trying to determine a more fundamental question --- whether the model itself is a good one. Parameter fitting can reveal weaknesses in a model when the model is unable to reproduce the measurements.  However, it says little about whether that model outperforms another model when both are good fits.  

The discovery of the accelerating universe was not technically a discovery of acceleration, but rather a discovery that a model that included cosmological-constant-like dark energy outperformed a model with only matter and radiation. \citep[Direct detection of the acceleration may be possible in the future by long-term monitoring of the redshift evolution of extra-galactic spectra, known as the `Sandage-Loeb' effect;][]{Sandage62,Loeb98}.

Prior to the supernova data, the preferred model for our universe was general relativity with cold dark matter (CDM).  To calculate the expansion history in such a model you only need to know the matter density\index{matter density} $\om$ and the Hubble constant\index{Hubble constant} $H_0$, often written as $h=H_0/100$ \kmsmpc.  Type Ia supernovae proved difficult to fit with such a model.  By adding one parameter, i.e.\ a cosmological constant $\oll$, the new model ($\Lambda$CDM) was a much better fit.  In the $\Lambda$CDM model the universe accelerates.  Thus the discovery of ``acceleration''. 

However, a model with an extra parameter will always be able to outperform a model that has a subset of its parameters, because the extra flexibility can only improve the fit.  Whether the extra parameter is justified requires some assessment of model selection -- i.e.\ is the more complex model enough of a better fit to justify the addition of the extra parameter?  In the case of $\oll$ the answer was a most resounding ``yes!".  

Here we will briefly summarise the methods by which such statements are quantified, both parameter fitting and model selection.  Far more detail on all of these basics can be found in \citet{davisparkinson16_stats}.

\subsection{Parameter fitting}

In Bayesian statistics the parameter likelihood $\mathcal{L}(\{\theta\})=P(D|\{\theta\})$ is the probability of the data (the `results') given some parameter values ($\{\theta\}$, the `causes').  What we are usually trying to figure out is the reverse of this ---  the likelihood\index{likelihood} of the parameter values given the data --- which is known as the posterior $P(\{\theta\}|D)$.  The two are not exactly the same, but are related to each other through Bayes' theorem, which incorporates any prior knowledge that should be included in the calculation $P(\{\theta\})$,
\beq P(\{\theta\}|D) \propto P(D|\{\theta\})P(\{\theta\})\,. \label{eqn:bayes}\eeq
For data with a large number of measurements (such as supernovae, where a large number of photons are incident on the detector), the likelihood function $P(D|\{\theta\})$ will converge to a Gaussian\index{Gaussian}, through the central limit theorem. In this case the likelihood can be evaluated through a $\chi^2$ calculation, since the $\chi^2$ is proportional to the log of the likelihood,  
\beq \mathcal{L} = e^{-\chi^2/2} \label{eqn:chi2}. \eeq
The value of $\chi^2$ in the simplest case of independent data points for a particular data/model combination is given by,
\beq 
\label{eqn:chi2}
\chi^2 = \sum_i\left(\frac{\mu_{{\rm model},i} - \mu_{i}}{\sigma_{i}}\right)^2,
\eeq
where $\mu_{{\rm model},i}$ is the value for data point $i$ predicted by the model and $\mu_i\pm\sigma_i$ is the $i$th data point and uncertainty. 
We then evaluate the value of $\chi^2$ at every point in the parameter space $\{\theta\}$ of the model being tested.  
Once the $\chi^2$ has been computed across the range of parameter values, it can be used to determine the likelihood $\mathcal{L}(\{\theta\})$. Then the prior can be applied to determine the posterior probability distribution of the parameters, given the data.

There are many different approaches to sampling the parameter space and generating the posterior statistics. For example, a grid search can become computationally expensive and inefficient as the dimension of the search space becomes large, and so adaptive or Monte-Carlo methods are used instead. For a summary of some of these methods, we refer to a review by the authors \citep{davisparkinson16_stats}.

\subsection{Model testing}

Finding the best fit parameters in a model is not enough, one should also always test whether the best fit is a good fit.  The reduced $\chi^2$ is a simple way to get a feel for which model best describes the data but alone is not enough.   
While it gives a rough estimate of whether a particular model is a good fit to the data it does poorly at comparing models when those models have differing numbers of free parameters.  For example, adding a cosmological constant to the CDM model to make it $\Lambda$CDM gave the model extra flexibility, so it is guaranteed to be a better fit than CDM alone.  Thus a lower $\chi^2/$dof (where dof is the number of degrees of freedom in fitting the data, which can normally be evaluated as the number of independent data points minus the number of model parameters) is not a good enough criterion on its own to judge the relative merits of a model. 

{\bf Bayesian Evidence}\index{Bayesian Evidence} quantifies how well a model describes the data weighted by the amount of parameter space it could have covered.  A simpler model has less parameter space, so if it is able to fit the data as well as a more complex model then it will be preferred by the Bayesian Evidence calculation.  In essence this is quantifying Occam's razor.  An example of the use of the Bayesian evidence to choose between parameterisations for the dark energy equation of state is given in \cite{DEEvidence}.

{\bf Information Criteria}\index{Information Criteria} (IC) are another method to penalise more complex models.  They simply ask whether a parameter is worthwhile adding to a model by comparing the best-fit $\chi^2$ before and after the parameter has been added.  There are several options including the Akaike IC and the Bayesian IC, which differ slightly in their level of penalty for adding extra parameters \citep{liddle1997}.   An example of these IC applied to supernova cosmology can be found in \cite{davis07}.

\section{The variety of models to be tested}
\label{sec:variety}\label{sect:models}

Models of dark energy take many forms.  After establishing the basics of the standard model (Sect.~\ref{sect:standard}) we begin by describing {\em parametric models} that consider dark energy time evolution following particular functional forms (Sect.~\ref{sec:para}) and then consider {\em non-parametric models} that allow more flexible time-varying dark energy (Sect.~\ref{sec:non-para}).  These are both methods driven by observation, where the functional forms and methods are motivated by the desire to generically test whether dark energy varies with time, or remains consistent with a cosmological constant.  We then go on to describe models driven by theory, such as dark fluid models, $f(R)$ gravity, and braneworld models (Sect.~\ref{sec:beyond} and \ref{sec:modified}).  These make more specific predictions for how dark energy could possibly vary with time or position. 

\subsection{The standard $\Lambda$CDM model}\label{sect:standard}

The standard cosmological model is an on-average homogeneous, isotropic universe that is spatially flat, consisting of matter, radiation, and a cosmological constant.  For late-time measurements the radiation is negligible, so the contents of the universe are completely designated by a single parameter, the matter density $\om = \rho_{m}/\rho_{\rm crit}$, which is defined relative to the critical density $\rho_{\rm crit}$, and includes both luminous matter and cold (non-relativistic) dark matter (CDM).  Since the universe in this model is flat the cosmological constant is simply given by $\oll=1-\om$.  

This standard $\Lambda$CDM model belongs to a more general family of models with non-zero curvature, characterized by the Friedmann-Robertson-Walker (FRW) metric\index{Friedmann-Robertson-Walker metric}, which describes spacetime in an on-average homogeneous and isotropic universe. The metric is given by 
\beq {\rm d}s^2 = -c^2 {\rm d}t^2 + R^2(t) \left[{\rm d}\chi^2 +S_k^2(\chi)({\rm d}\theta^2+\sin^2\theta\, {\rm d}\phi^2)\right].\eeq
Here $t$ is cosmological time, $\chi$ is comoving distance, while $\theta$ and $\phi$ are the angular position in spherical coordinates. The scale factor $R(t)$ is often normalised to its value at the present day $a(t)=R(t)/R_0$, and $S_k(\chi) = \sin(\chi), \chi, \sinh(\chi)$ in closed, flat, and open geometries respectively.  The Hubble parameter is defined as the ratio of the rate of change of scale factor to the scale factor itself, $H(t)=\dot{R}/R=\dot{a}/a$, where overdot represents differentiation with respect to time.  More often $H(t)$ is replaced with $H(z)$ since it is computationally and observationally easier to represent time by the redshift of a galaxy that emitted the light we now see at that time.  Comoving distance is,
\beq d(\bar{z}) = R_0\chi(\bar{z}) = c\int_0^{\bar{z}}\frac{dz}{H(z)},  \eeq
where $\bar{z}$ is the cosmological redshift (with no contribution from peculiar velocities or lensing).  Luminosity distance, which is the distance measured by supernovae, is related to comoving distance by,
\beq d_L(z) = (1+z) R_0 S_k\left(\chi(\bar{z})\right).\eeq 
Note the two different redshifts in this equation.  The comoving distance\index{comoving distance} is governed by the cosmological redshift\index{cosmological redshift}, but the pre-factor (that arises due to dilution of photon number counts and beaming) is sensitive to the observed redshift.  Using the cosmological redshift in both terms gives negligible error at the level of current experiments, but using the observed redshift in both gives significant errors \citep{calcino_thesis2015}.

The standard cosmological model,  $\Lambda$CDM, is a universe that is spatially flat, consisting of matter, radiation, and a cosmological constant.  For late-time measurements the radiation is negligible, so the contents of the universe are completely designated by a single parameter, the matter density $\om = \rho_{m}/\rho_{\rm crit}$, which is defined relative to the critical density $\rho_{\rm crit}$.  Since the universe in this model is flat the cosmological constant is simply given by $\oll=1-\om$.  

By inserting the stress-energy-momentum equation relevant for an isotropic, homogeneous, perfect fluid, into Einstein's equations, one can derive Friedmann's equation\index{Friedmann's equation}, which governs the dynamics of the expansion.  Taking into account many contributions to the energy density of the universe ($\Omega_i$) the Friedmann equation is concisely expressed as,
\beq H^2(z) = H_0^2 \sum_i \Omega_i (1+z)^{3(1+w_i)}, \label{eqn:Friedmann} \eeq
where the equation of state parameter, $w=p/\rho$, is the ratio of pressure to density of each component ($w=\frac{1}{3},0,-1$ for radiation, matter, and cosmological constant respectively).  In the $\Lambda$CDM model that becomes $H(z)=H_0[\om a^{-3} + \ok a^{-2}+\oll]^{1/2}$.  Note that in non-flat universes it is important to include a curvature term $\ok=1-\sum \Omega_i$, with $w_k=-\frac{1}{3}$.  With these basics established, we now move on to extensions and variations on this model.

\subsection{Parametric Models, simple extensions to $\Lambda$CDM}
\label{sec:para}

The simplest alternative to a cosmological constant, where the energy density is constant with time, would be some time-varying dark energy. Without reference to any physical model, this is normally parameterised by $w_{\rm DE}$, which characterises the equation of state\index{equation of state} of dark energy (DE).  For an arbitrary equation of state, where $w_{\rm DE}$ can take any value as a function of time or scale factor, the energy density can be found by solving the continuity equation\index{continuity equation}, to give
\begin{equation}
\rho_{\rm DE}(\bar{z}) = \rho^0_{\rm DE}\exp\left\{\int_0^{\bar{z}} \frac{3(1+w_{\rm DE}(z))}{1+z}{\rm d} z\right\}\,.
\end{equation}
If the equation of state is constant with redshift, then this simply becomes,
\begin{equation}
\rho_{\rm DE}(\bar{z}) = \rho^0_{\rm DE}(1+\bar{z})^{3(1+w_{\rm DE})}\,.
\end{equation}
The cosmological constant is the only model that predicts an equation of state $w_{\Lambda}=-1$ for all epochs, though we will see later that some models can predict an equation of state very close to that of $\Lambda$ at late times (often by design!). Because of this, any decisive detection of $w_{\rm DE} \ne -1$ at any epoch would automatically rule out the cosmological constant as a convincing explanation for the acceleration.

Current constraints place $w_{\rm DE}$ very close to $-1$. A data compilation that includes Cosmic Microwave Background temperature and polarisation data from the Planck surveyor, Baryon Acoustic Oscillation data from 6dFGS, SDSS-MGS, BOSS-LOWZ and CMASS-DR11, supernovae data from the Joint Lightcurve Analysis, and  $H_0$ measurements from Cepheids,  gives a combined constraint on a constant equation of state of $w = -1.019^{+0.075}_{-0.080}$ at 95\% confidence \citep{Planckcosmoparams15}. The modern challenge for determining the dark energy equation of state is not determining the value today, which seems to be very close to $-1$ indeed, but rather determining if it was different to $-1$ in the past.

\begin{table}[th]
\centering 
\renewcommand{\arraystretch}{1.5}
\begin{tabular}{lcl}
{\bf Parameterisation} & & {\bf Reference} \\ \hline
\multirow{ 2}{*}{Linear in redshift} &  \multirow{ 2}{*}{ $w(z) = w_0 + w_1z$} & \citet{hutererturner1999},\\
& &  \citet{welleralbrecht(2001)} \\ \hline
\multirow{ 2}{*}{Linear in scale factor} & \multirow{ 2}{*}{ 
$\begin{array} {lcl} w(a) & = & w_0 + w_a(1-a) \\ & = &w_0 + w_1\frac{z}{1+z} \end{array}$}
   &  \citet{chevallierpolarski(2001)}, \\ 
   & & \citet{linder03}  \\   \hline
Nonlinear in scale-factor &  $ w(a) =  w_0 + w_a\frac{z}{(1+z)^2} $ & \citet{Jassal2005} \\ \hline
Logarithmic in redshift & $w(z) = w_0 + w_1\ln(1+z) $ & \citet{efstathiou2000} \\ \hline
 \multirow{ 3}{*}{ A step-like function} &  \multirow{ 3}{*}{ $\begin{array}{lcl} w(z) & = & w_0 + \frac{w_\infty - w_0}{1+\exp[(z-z_t)/\Delta]}\\ & = &  w_a + (w_0 - w_a )\frac{a^{1/\tau} [1-(a/a_t)^{1/\tau}]}{1-a^{-1/\tau}}\end{array} $} & \citet{corasaniti2004}, \\
 & & \citet{hannestad04}, \\
& &   \citet{defelice2012}\\ \hline
Hybrid model & $ w(z)=\left\{ 
\begin{array}{lll}
w_{0}+w_{1}z &~~{\rm if}~~& z<1 \\ 
w_{0}+w_{1} &~~{\rm if}~~& z \ge 1.
\end{array}\right .
$ & \citet{upadhye2005} \\\hline
\end{tabular}
\caption{\label{tab:wde}A number of different parameterisations for the scale-factor or redshift-dependence of the dark energy equation of state.}
\end{table}

A number of different parameterisations for a time-varying equation of state have been proposed. These are shown in Table \ref{tab:wde}.

The utility of these to actually describe some physical mechanism that drives the acceleration, has been discussed in the literature \citep[e.g.][]{bassett2003,corasaniti2004}. The models linear in redshift or scale-factor may describe some late-time dynamics of the dark energy very well, but have difficulty matching the conditions of the Universe at recombination or big bang nucleosynthesis, for example. The step function on the other hand, may describe some tracking quintessence model, which tracks the dynamics of the early Universe (such that $w_\infty = 1/3$ or 0), but later transitions to drive the acceleration at late time (i.e. $w_0 \rightarrow -1$). We will discuss these more physically motivated models further in section \ref{sec:beyond}.

Another simple extension, or rather alternative to the standard $\Lambda$CDM cosmology would be a different choice of metric, where we relax the assumption of an homogeneous universe.  An apparent acceleration can be generated if we live in a vast under-density, or void.  Such models are described by the Lema\^{i}tre-Tolman-Bondi metric, and although they can explain the supernova data, they fail other observational tests (as well as the Copernican principle).   

The alternative to an apparent acceleration generated by a void would be a real acceleration, but generated without recourse to new physics. In these `backreaction' models, the non-linear nature of general relativity would generate acceleration or apparent acceleration as a consequence of the inhomogeneities generated by structure formation \citep[e.g][]{Rasanen2004,wiltshire07,wiltshire09,KolbMatarese2010}.

\subsection{Non-parametric Models}
\label{sec:non-para} 

The simple model of parametrised ignorance above, described by the equation of state $w$ and the different parameterisations for the time-variation, can be limited in usefulness. Such models may not adequately or successfully describe some features present in the data. A non-parametric approach, where the analysis is driven by the data rather than the initial model, will always be more successful in this area, with the caveat that they are often more difficult to connect to the underlying physics of what is occurring. We consider two such cases here.

\subsubsection{Cosmographic parameters}
\label{sec:cosmographic}

In the cosmographic approach  the connection between the dynamical history of the Universe and its material components (i.e. the Einstein equations) is discarded in favour of direct reconstruction of the kinematical history of the expansion \citep{Visser2004,Cattoen2008,Capozziello2011,Lazkoz:2013by}. The analysis proceeds in terms of the cosmographic parameters\index{cosmographic parameters}, which are the Hubble rate $H(t)$, the deceleration parameter $q(t)$, and the jerk $j(t)$, snap $s(t)$, and lerk $l(t)$ parameters. These are defined to be,
\beq 
\label{eq:qParam}
F_n = \frac{1}{a}\frac{d^na}{dt^n} \left[\frac{1}{a}\frac{da}{dt}\right]^{-n} 
\eeq
with $H(t) = F_0$, $q(t)=-F_1$, $j(t)=F_2$ etc.  We construct the Taylor series of the expansion history, using the current values of cosmographic parameters (where subscript $0$ indicates the value at the present day $q_0 \equiv q(t_0)$ etc.),
\bea
\frac{a(t)}{a(t_0)} & =&  1+ H_0(t-t_0) - \frac{q_0}{2}H_0^2(t-t_0)^2 + \frac{j_0}{3!}H_0^3(t-t_0)^3  \nonumber\\
&& + \frac{s_0}{4!}H^4_0(t-t_0)^4 + \frac{l_0}{5!}H^5_0(t-t_0)^5 + O[(t-t_0)^6].
\label{eq:scaleFactor}
\eea
The expansion history can therefore be used to make predictions regarding the physical distance travelled by a photon, and so the measured luminosity distances can be used to make inferences about the values of the cosmographic parameters.

\subsubsection{Reconstruction of the equation of state}

The equation of state remains our best diagnostic for a physical explanation for the dark energy beyond the cosmological constant. Rather than a parameterised approach, many have suggested we reconstruct the values of $w(z)$ through non-parametric approaches. Some of these involve manipulating or smoothing the luminosity distance data from supernovae, and then computing the reconstructed  Hubble rate $H(z)$ (or expansion rate $E(z)=H(z)/H_0$, since the Hubble rate today $H_0$ is often a nuisance parameter) and equation of state $w(z)$. Methods that have been used include (but are not limited to):
\begin{itemize}
\item Polynomial fits to the luminosity distance \citep{dalydjorgovski2003,dalydjorgovski2004,daly2008} 
\item Smoothing the data using Gaussian kernels \citep{shafieloo2006}
\item Reconstruction using the Bayesian Maximum Entropy method \citep{zunckel2007}
\item Reconstruction using Gaussian Processes \citep{holsclaw2010}
\end{itemize}

All of the above methods are effectively doing the same thing, regularising or reconstructing the stochastic luminosity distance data to give the best possible estimate of the distances to different redshifts, and then taking the first and second derivatives to estimate the equation of state. 
An alternative reconstruction approach allows the largest possible freedom of $w(z)$ in a binned approach, letting the data choose the widths and positions of the redshift bins \citep[e.g.][]{sollerman09}. The best choice of redshift bins are those  that decorrelate the covariances in the parameter errors between the bins (de Putter and Linder 2008). A principal component analysis will determine the eigenmodes for which this is true, though these eigenmodes will vary widely for different data sets.

\subsection{Alternative physical explanations} 
\label{sec:beyond}

We now turn our attention to directly testing alternatives to the simple cosmological constant $\Lambda$. We divide these other explanations for the acceleration roughly into two categories.  We first discuss those that propose some extra field or fluid that directly drives the acceleration, then cover those where the Einstein equations of gravity are modified to give rise to a late-time acceleration.

The possibility of having some previously unknown physics driving the acceleration can be quantified by defining some extra degree of freedom (which is in most cases a scalar with some evolution equation) and feeding its effect back into the Friedmann equation. Thus supernovae and other distance measurements can be used to constrain the evolution of that new scalar, and determine the free parameters associated with that model.

\subsubsection{Dark fluid models}
\label{sec:darkfluid}

The earliest (and one of the simplest) dark fluids is a minimally-coupled, light scalar field, also known as {\em quintessence}\index{quintessence} \citep{ratra1988,wetterich1988}. The evolution equation for a scalar field is the Klein-Gordon equation,
\begin{equation}
\Box\phi = -\frac{\d V(\phi)}{\d\phi}\,,
\end{equation}
where $\phi$ is the quintessence field, $V(\phi)$ is the potential governing the evolution of the field, and $\Box$ is the d'Alembertian operator. In a FLRW universe, the d'Alembertian contains a term relating the rate of expansion, slowing the field evolution (a `Hubble friction' term), and so the Friedman equation feeds into the Klein-Gordon equation. Similarly the energy density of the field feeds into the Friedman equation, driving the acceleration. In a homogeneous universe, the spatial-dependence of the field can be neglected, and so the energy density ($\rho_\phi$) and pressure  ($p_\phi$) can be written purely in terms of the background, time-varying field $\phi(t)$ as
\begin{eqnarray}
\label{eqn:quint}
\rho_{\phi} = \frac{\dot{\phi}(t)^2}{2} + V(\phi) & \,,~~~ & p_{\phi} =  \frac{\dot{\phi}(t)^2}{2} - V(\phi)\,.
\end{eqnarray}
 The equation of state $w_\phi=p_\phi/\rho_\phi$ is bounded $-1 < w_\phi < 1$, complying with the weak energy condition from General Relativity. It will tend towards $-1$ in the potential-dominated regime, where $V(\phi) \gg \dot{\phi}^2$. Once the potential and the initial conditions for the field have been specified, the Klein-Gordon equation for the background field $\phi(t)$ and Friedmann equation must then be solved simultaneously.

The field is normally specified to be `light' such that the gradient and curvature of the potential are small, and so the field remains potential dominated over the required timescales. Quintessence models can be roughly  divided further into `thawing' models, in which the field value is  frozen by the Hubble friction in the early cosmological epoch and so are initially potential-dominated and only start to evolve at late times, and `freezing' models, where the dynamics of the scalar can track or scale with the other components (radiation or matter), and only become potential dominated at very late times. However, since the excursion of the field can still be large (of order the Planck scale), and the degree of fine-tuning required to give the observed acceleration is also large, such simple quintessence models still have the same issues that plague the cosmological constant. A review of the current state of quintessence is given by \citet{tsujikawa2013}.

Beyond standard quintessence there are a host of other models, including (but not limited to): 
\begin{description}
\item [{\em k-essence models},]{with a non-standard kinetic term in the action \citep{armendarizpicon(2001)}}.
\item [{\em Phantom models},]{where the weak energy condition is violated and the equation of state can be $w<-1$ \citep{caldwell2002}}.
\item [{\em Coupled dark energy models},] where the coincidence problem is alleviated by positing some relation between the dark energy and the matter or neutrino component \citep{amendola2000}.
\item  [{\em Chameleon scalar fields},] where the strong coupling that is expected between scalar and matter fields is shielded by the chameleon mechanism \citep{khouryweltman2004}.
\item  [{\em Unified dark energy-dark matter models},] where both phenomena are aspects of a single fluid, such as the Chaplygin gas models \citep{Kamenshchik(2001)}.
\end{description}
 
\subsection{Models that alter gravity}
\label{sec:modified}

Modified (or alternative) gravity theories suggest that the late-time acceleration is a natural consequence of the theory of gravity being different to  Einstein's theory. 
These theories may still contain the cosmological constant, but go on to suggest a more natural mechanism for the observed value, thus bringing it in line with the prediction from quantum mechanics. We summarise the more commonly discussed models here.

(We note that many of these theories are often described as making different predictions from that of General Relativity.  However they do not, since all General Relativity requires is coordinate independent physics, which remains true for the majority of alternative gravity theories. Instead these theories make different predictions to {\em Einsteinian gravity}, which assumes the standard Einstein equations.)

\subsubsection{$f(R)$ gravity}

The field equations in general relativity are usually derived using the principle of least action.  The gravitational part of the {\em Einstein-Hilbert action}\index{Einstein-Hilbert action} is given by 
\beq S=\frac{1}{16\pi G}\int {\rm d}^4x\sqrt{-g} \,R\, \, \eeq
where $G$ is the  gravitational constant, and $g$ is the determinant of the metric $g_{\mu\nu}$. We work in units where the speed of light $c=1$.
In $f(R)$ gravity the standard action is allowed to be one step more complex, by replacing the Ricci scalar, $R$, with an unspecified function of the Ricci scalar, $f(R)$. Until recently our experience with gravity has been in relatively dense environments, such as the solar system or the galaxy, with relatively large curvature (Ricci scalar). It therefore may be that in the empty voids between galaxies, where spacetime curvature becomes very small, that extra correction terms need to be applied to the action, which could give rise to acceleration,
\begin{equation}
S = \frac{1}{16\pi G} \int {\rm d}^4 x \sqrt{-g} \,f(R).
\end{equation}
The exact function is freely chosen, though if $f(R) = R - 2\Lambda$, we recover the usual Einstein equations with a cosmological constant. 
 
We usually obtain Einstein's field equations by minimising the action, which, for the standard Einstein-Hilbert action, gives,
\beq R_{\mu\nu} - \frac{1}{2} g_{\mu\nu}R = 8\pi G T_{\mu\nu}, \label{eq:einstein} \eeq
where $T_{\mu\nu}$ is the standard stress-energy tensor.  By inserting the $T_{\mu\nu}$ for a perfect fluid into eqn.~\ref{eq:einstein} the Friedmann equation (eqn.~\ref{eqn:Friedmann}) follows. Once $f(R)$ is introduced the more general derivation gives, 
\begin{equation}
f_{R}R_{\mu\nu}  -\nabla_\mu \nabla_\nu f_{R} - \left(\frac{f(R)}{2} - \Box f_{R}\right)g_{\mu\nu} = 8\pi G T_{\mu\nu}\,,
\label{eq:Varying}
\end{equation}
where $f_{R}=\partial f(R)/\partial R$ and $\Box\equiv\nabla^\alpha\nabla_\alpha$. Once the function $f(R)$ is chosen, then the Friedmann-Robertson-Walker metric, Ricci scalar, and Ricci tensor can be inserted into eqn.~\ref{eq:Varying}, and the modified Friedmann equation can be derived.

In order to choose a viable $f(R)$ function, there are a number of stability conditions that must be considered, including {\bf no anti-gravity:} which holds that $f_{'R}>0$ for $R\ge R_0$, where $R_0$ is the present value of the Ricci scalar; {\bf consistency with local gravity tests:} where the second derivative $f_{'RR}>0$ for $R\ge R_0$, and $f(R) \rightarrow R - 2\Lambda$ for $R\gg R_0$; and {\bf stability of the late-time de-Sitter point:} where $0 < \frac{Rf_{'RR}}{f_{'R}} < 1$ at $r = -\frac{Rf_{'R}}{f} = -2$  \citep{Amendola2007}.

\begin{table}[ht]
\centering 
\renewcommand{\arraystretch}{2.5}
\begin{tabular}{lclcl}
${\mathbf f(R)}$ {\bf function} & & & & {\bf Reference} \\ \hline
$f(R) = R - R_0 \frac{c_1(-R/R_0)^n}{c_2(R/R_0)^n-1} $&~~~& $R_0,~ n,~ c_1~\&~c_2 >0$ &~~~& \citet{HuSawicki2007} \\ \hline
$f(R) = R - \lambda R_{0} \left[1-\left(1+\frac{R^2}{R_{0}^2} \right)^{-n}\right]$ &~~~& $R_0,~n~\&~\lambda >0$ &~~~& \citet{Starobinsky2007} \\ \hline
$f(R) = R - \lambda R_0\left(\frac{R}{R_0}\right)^n$ &~~~& $0 < n < 1$, $R_0~\&~\lambda >0$ &~~~& \citet{Amendola2007}\\ \hline
 \multirow{ 2}{*}{$f(R) = R - \lambda R_0\tanh \left(\frac{R}{R_0}\right)$} &~~~&  \multirow{ 2}{*}{ $R_0~\&~\lambda>0$}  &~~~&\citet{ApplebyBattye2007}, \\
&~~~&  &~~~& \citet{Tsujikawa2008} \\ \hline
\end{tabular}
\caption{\label{tab:fR} Various functions of the Ricci scalar that can give rise to a late time acceleration, and the stability conditions that are required \citep[from][]{AmendolaTsujikawaBook}.}
\end{table}

A list of some of the proposed $f(R)$ models that satisfy these conditions is given in Table~\ref{tab:fR}. All of these models contain a threshold value of the Ricci scalar $R_0$, which has to be tuned to the correct value in order to manifest the observed late-time acceleration. So these models also suffer from the fine-tuning problem of the cosmological constant. In a sense, all of these models have been `designed' to recover the observed acceleration. However, they should not be considered as final theories, but rather as `effective theories', that phenomenologically describe some undiscovered new explanation or theory of gravity.

\subsubsection{Braneworld models}
\label{sec:brane}

The idea of `branes'\index{branes} (short for membranes) as lower dimensional objects in a higher dimensional space first arose out of string theory. In Kaluza-Klein\index{Kaluza-Klein} theory, all extra dimensions are compactified, but in braneworld models, a 3(spatial)-dimensional brane is embedded in a 5D bulk with large extra dimensions. In these models standard model particles are open strings, whose ends must be fixed on a brane, whereas gravitons (which are closed strings) are allowed to propagate in the bulk. Thus it may be possible to generate acceleration by having gravity `leak' out of the brane into the bulk \citep{Deffayet2002}.

The DGP braneworld model proposed by \citet{dvali2000} consists of a 3-brane embedded in a Minkowski bulk spacetime with infinitely large extra dimensions. In the standard DGP model, 4D gravity (Newton's law) is recovered on small distances, whereas the effect for the 5D gravity manifests as a modification only on large distances. There is a solution to the DGP model that is `self-accelerating', requiring no extra dark fluid to generate the observed acceleration.
For a flat geometry, the modified Friedmann equation is given by,
\begin{equation}
H^2 - 2\epsilon\frac{G_{(4)}}{G_{(5)}} H = \frac{8\pi G_{(4)}}{3}\rho\,,
\end{equation}
where $G_{(4)}$ is the value of Newton's gravitational constant induced on the brane, whereas $G_{(5)}$ is the value of the gravitational constant in the higher dimensional bulk, and $\epsilon = \pm 1$, depending on how the scale-factor $a$ is changing with respect to the extra-dimensional coordinate distance. For $\epsilon=+1$, and a Universe dominated by non-relativistic matter, the expansion approaches a de Sitter (accelerating) solution,
\begin{equation}
H \rightarrow H_{\rm dS} = 2\frac{G_{(4)}}{G_{(5)}}.
\end{equation}
By tuning the value of the 5-dimensional gravitational constant $G_{(5)}$, such that the ratio is of the order of the present Hubble radius $H_0^{-1}$, we recover a late-time acceleration on the correct time-scale to fit the observed value.

This particular braneworld model seems like the most `natural' model to explain the acceleration, as it requires no value of the cosmological constant to be inserted by hand, and allows us to learn something about extra dimensions that is beyond the ability of current particle accelerators. However, the effective value of the equation of state today, $w_0$, is required in this theory to be approximately \citep{MaartensMajerotto2006},
\begin{equation}
w_{\rm eff} = -\frac{1}{1+\Omega_m}.
\end{equation}
For a value of $\Omega_m \simeq 0.3$, this gives us $w_{\rm eff} \simeq -0.77$, which is excluded by current data. Furthermore, a perturbation theory analysis shows that the original DGP model contains a ghost instability\index{ghost instability} -- that is a term with negative kinetic energy that creates a vacuum instability \citep{Gorbunov2006}. Thus the DGP model is ruled out by both theory and experiment.

\subsubsection{Galileon Models}

\index{Galileon}Galileon scalar fields are those that are invariant under a shift symmetry in field space, $\partial_\mu \pi \rightarrow \partial_\mu \pi + c_\mu$, where $\pi_\mu$ is the scalar field. These theories can be considered as fully covariant forms of the DGP theory (see the previous section, \ref{sec:brane}), and because of their symmetry, have the advantage that the $\pi$ field equations will be at most second order \citep{Deffayet2009}. In fact there are only five possible Lagrangians for the Galileon that remain second order when coupled to gravity, and do not introduce ghost instabilities \citep{Nicolis2009}.

The Galileon scalar contains a natural screening mechanism, the Vainshtein effect\index{Vainshtein effect}, where non-linear effects decouple it from the matter sector. Thus the Galileon has no effect on the expansion rate during nucleosynthesis and recombination, and only becomes important at late times \citep{Chow2009}. At late-times the stable, self-accelerating branch has a very negative equation of state, $w_\pi < -1$. While it is possible to find regions of the Galileon parameter space that match the measured background history, there is then a tension between predictions for and measurements of the growth of structure \citep{ApplebyLinder2012}.

\section{Beyond the Hubble Diagram}
\label{sec:beyondhubble}

If the expansion history remains consistent with $\Lambda$CDM, then there is no way even in principle to use the Hubble diagram\index{Hubble diagram} to distinguish between many models of dark energy because they can all reproduce the $\Lambda$CDM expansion (as they are designed to do).   However, structure formation can differ even in models that predict the same expansion history.  Therefore we look to the growth of structure, and lensing of light around that structure, to find signals that distinguish between models. 
  
In order to model the formation of structure, and the distortion of light by these structures, we need a metric that can model inhomogeneities. Using the conformal Newtonian gauge\index{conformal Newtonian gauge} we can consider perturbations to the Friedman-Robertson-Walker metric, a temporal perturbation given by $\psi$ and a spatial perturbation given by $\Phi$ such that, 
\begin{equation}
 ds^2 = a^2\left[-(1+2\Psi)d\tau^2+(1-2\Phi)d\bar{x}^2\right],
 \end{equation}
where $a$ is the scale factor, $\tau$ conformal time, and $\bar{x}$ is the vector of spatial coordinates. 
In general relativity (and in the absence of any large-scale anisotropic stress) $\Psi=\Phi$, so there is no distinction between the spatial and temporal metric perturbations.  However, other theories allow more flexibility, and this is a particularly interesting split because different measurable phenomena behave differently in the two components.  For example lensing is sensitive to both the time and space components, but clustering of matter is purely spatial.  

Supernovae can trace large scale structure, both because they are good distance indicators (useful for peculiar velocities) and because they are good standard candles (useful for lensing magnification).  Peculiar velocities\index{Peculiar velocities} dominate the dispersion at low redshifts ($z\lsim 0.3$), whereas lensing dominates at higher redshifts.  As data quality improves both of these effects need to be taken into account to achieve unbiased cosmological constraints from the Hubble diagram.  More interestingly, though, they can be used as a signal in their own right to distinguish between different theories explaining dark energy and dark matter.  In this section we discuss how supernovae can be used to make these two different types of measurements.

\subsection{Peculiar Velocities}

Peculiar velocities, like the acceleration, cannot be directly measured, since all that can be taken from a galaxy spectrum is the total redshift. They are inferred by comparing a precise measurement of distance (for example, using a supernova as a standard candle) to the distance expected given the object's redshift (though in practice, we usually do not bother to convert the observables to distances, but work directly in terms of the discrepancies in magnitude). In a homogeneous universe, where all objects sit in the Hubble flow, there would be no difference, but in an inhomogeneous universe discrepancies  arise due to an extra redshift (or blueshift) created by local motion.   Denoting the cosmological redshift, $\bar{z}$, and the peculiar velocity redshift, $z_{\rm p}$, the observed redshift, $z$, is given by,
\beq (1+z) = (1+\bar{z})(1+z_{\rm p}). \eeq
 Traditionally supernova cosmology analyses that fit the magnitude-redshift relation exclude or down-weight nearby supernovae at $z<0.02$, because closer than that the contribution from peculiar velocities (which are $\sim300~\kms$) exceeds 5\% and due to correlations between supernova motions the error can be systematic (particularly for surveys covering small regions of sky). 

Peculiar velocity measurements are often made using galactic distance measures like Tully-Fisher\index{Tully-Fisher} \citep{TullyFisher} and Fundamental Plane\index{Fundamental Plane} \citep{Djorgovski1987,Dressler1987}.  While these methods have the advantage of large numbers, the errors associated with them are often large and very non-Gaussian, leading to biases that need to be carefully accounted for \citep{johnson14,scrimgeour16}. Supernovae on the other hand have much tighter and more Gaussian probability distributions, making them excellent tools for peculiar velocities. 

Velocity perturbations are generated by density perturbations through the gradient of the gravitational potential $\nabla \phi$. Since the distribution of matter can be modelled/measured as a power spectrum, the distribution of velocities can also be described using a power spectrum. Peculiar velocities are particularly useful because the motion of galaxies is induced by the entire density field, including dark matter, so they are not limited in the manner of galaxy surveys, which only directly trace the density of {\em visible} matter. Moreover, peculiar velocities are sensitive to very large scale density fluctuations\index{density fluctuations}, beyond the scale that is easy testable in galaxy redshift surveys that measure clustering \citep{johnson14}.  By measuring the power spectrum of galaxy motions, you get a direct measurement of how quickly structures are growing.  This is a useful measure to observationally distinguish between different theories \citep{jennings12,koda14}.

There are several ways in which peculiar velocities can be compared to models. Peculiar velocities manifest themselves as a scatter in the Hubble diagram, but the scatter is not random.  This could be a source of systematic error, since supernovae that are part of the same velocity field will have  related offsets in redshift. 
Using a model power spectrum one can predict the strength of correlations in supernova magnitudes \citep{hui06,davis11}.  This allows you to remove the impact of peculiar velocities on the magnitude-redshift cosmology estimates by using the covariance matrix to down-weight the supernovae that should have correlated velocities 
\citep[the detailed equations can be found in][]{davis11}.

However, if you want to compare the strength of the correlations to theory, and so use the size of the velocities as a cosmological probes, it is possible to calculate the covariance matrix generated by the peculiar velocities for various models, and find the one that best fits the data.  This is almost reversing what is usually done in Eq.~\ref{eqn:chi2}, because the model does not go into the distances, but into the covariance matrix, e.g.  
\beq {\mathcal L} = \frac{1}{| 2\pi C^{(v)}|^{1/2}}\exp \left(-\frac{1}{2} \sum_{ij} v_i({\bar x}) \, C^{(v)-1}_{ij} v_j({\bar x})\right). \eeq
Here the measured values of the velocity along the line of sight, $v$, at positions, ${\bar x}$, are being compared to the predicted correlations between them, as quantified by the inverse of the velocity covariance matrix for those positions $C^{(v)}$.  For details on how to calculate the covariance matrix for different theoretical models see \citet{johnson14}. 

Yet another method is to compare measured peculiar velocities to the predicted peculiar velocities inferred from maps of the density of galaxies \citep[e.g.][]{springob14,springob16,carrick15}.  This has been successful but encounters limitations because the fluctuation modes that affect velocities are often larger than encapsulated in the density data.

\subsection{Lensing}

It has long been known that supernova light is magnified and de-magnified as it rides the rollercoaster of curved space en route from the supernova to our telescopes \citep{frieman96,holz98,dodelson06}.  Dense lines of sight cause light to converge and result in magnification\index{magnification}, while emptier lines of site cause light to diverge and demagnify.  The strength of matter clustering as a function of redshift is sensitive to the cosmological parameters (particularly the matter density) and to the law of gravity.  Therefore measuring the lensing distribution can be used to measure cosmological parameters and test for deviations from standard gravity \citep[e.g. see][Fig.~5 for a plot of magnification probability for various cosmologies]{wang99}. 

Since clusters and filaments take up a relatively small volume compared to the voids between them, there are more under-dense lines of sight than over-dense ones.  So the signature of lensing is an asymmetric scatter about the Hubble diagram, with the median peaking on the faint side.  The signal increases with redshift, because the lines of sight are longer.  Typically we would expect additional magnitude dispersion due to lensing up to a maximum of $0.04$ mag for supernovae at $z<1$. 

This asymmetric distribution does not bias the Hubble diagram, since the number of photons must be conserved.  While many supernovae are slightly demagnified, a small number are highly magnified, and the mean is unaffected.  Biases only arise if the sample is incomplete (which can occur, for example, due to Malmquist bias\index{Malmquist bias} losing faint galaxies, or because highly lensed supernovae are obscured by the dust in the galaxies that would have magnified them).  

The magnification is related to the convergence $\kappa$ and can be estimated once we know the Hubble constant, the matter density, the total density, the distance to the supernova, and the density distribution along the line of sight. 
See Eq.~6.16 in \citet{bartelmann01} for details and \citet{smith14} for a practical implementation. 

Two ways of measuring the effect of lensing on supernova magnitudes are to:
\begin{enumerate}
\item Measure the asymmetry in the scatter about the Hubble diagram (fit to moments of the distribution),
\item Measure the cross-correlation between supernova magnitudes and foreground densities. 
\end{enumerate}

Detecting the signal of lensing requires many supernovae, so early attempts at both techniques gave varied results on the small supernova data sets available.  Measuring the moments in the distribution was attempted by \citet{wang05,castro14},  but careful modelling of the supernova magnitude measurements is required to rule out systematic error due to asymmetric observational uncertainties.
Cross-correlating supernova brightnesses with foreground densities has been performed by \citep{williams04,menard05,jonsson07,jonsson10,kronborg10,karpenka13}, and it is clear that modelling the mass distribution along the line of sight is important \citep[e.g.][]{mortsell01}.  The statistically most robust analysis to date (made using the largest sample of supernovae) was performed by \citet{smith14}, who found a positive but statistically marginal result (significance of 1.7$\sigma$).  

With upcoming surveys such as the Dark Energy Survey and the Large Synoptic Survey Telescope expecting to measure thousands of type Ia supernovae, the effect of lensing will be important to take into account for the standard Hubble diagram cosmological measurement, and will be strong enough to use these techniques to treat the lensing dispersion distribution as signal in its own right.

\newpage
\section{Conclusion}

This chapter reviewed some of the many varied models that try to explain the acceleration of the expansion of the universe (dark energy), and the ways in which we can test them.  Supernovae provide excellent measurements of the expansion history of the universe, but also can be used to measure the structure within it through measurements of peculiar velocities and gravitational lensing.  These latter features enable supernovae to distinguish between dark energy models that predict the same expansion history but differ in their predictions for growth of structure, in a way complementary to other large scale structure probes.  At the moment no theory to explain dark energy stands out as particularly compelling in relation to the others.  Therefore strong investment in theoretical efforts is required, and our observational efforts must keep looking for more precision and new ways to test dark energy, so we can provide compelling directions in which to guide theory.

\begin{acknowledgement}
We thank Edward Macaulay for helpful comments while this was under preparation. DP is supported by an Australian Research Council Future Fellowship [grant number FT130101086]. Parts of this research were conducted by the Australian Research Council Centre of Excellence for All-sky Astrophysics (CAASTRO), through project number CE110001020.

\end{acknowledgement}

\parindent = 0 mm







\printindex

\end{document}